\documentclass[a4paper]{article}
\usepackage{epsfig,amsmath,amsfonts,amssymb,setspace,multirow,textcomp}
\usepackage[T1]{fontenc}
\makeatletter
\renewcommand{\@evenfoot}{\hfil \thepage \hfil}
\renewcommand{\@oddfoot}{\hfil \thepage \hfil}
\makeatother

\renewenvironment{thebibliography}[1]{\begin{oldthebibliography}{#1}\setlength{\parskip}{0ex}\setlength{\itemsep}{0ex}}{\end{oldthebibliography}}

\begin{document}
\fontsize{11}{11}\selectfont 
\title{Meteor colorimetry with CMOS cameras}
\author{\textsl{B.\,E.~Zhilyaev, V.\,N.~Petukhov, V.\,N.~Reshetnyk, A.\,P.~Vid'machenko}}
\date{\vspace*{-6ex}}
\maketitle
\begin{center} {\small $Main Astronomical Observatory, NAS \,\, of Ukraine, Zabalotnoho \,27, 03680, Kyiv, Ukraine$}\\
{\tt zhilyaev@mao.kiev.ua}
\end{center}

\begin{abstract}
This article describes our approach to quantifying the characteristics of meteors such as temperature, chemical composition, and others. The program includes new algorithms for estimating temperature, heat radiation emitted by a fireball, and spectra of meteors containing emission lines. We are using a new approach based on colourimetry. We analyze an image of Leonid meteor-6230 obtained by Mike Hankey in 2012. Analysis of the temporal features of the meteoroid trail is performed. The main fluctuations in the brightness and wobbling of the meteor trail are observed at a frequency of about 3 Hz. The brightness variations in the integrated light are about 3\%. The amplitude of the wobbling is about 2\%. For determining the meteor characteristics we use the "tuning technique" in combination with a simulation model of intrusion. The progenitor of the meteor was found as an object weighing 900 kg at a speed of 36.5 km/s. The meteoroid reached a critical value of the pressure at an altitude of about 29 km in a time of about 4.6 sec with a residual mass of about 20 kg, and a residual speed of about 28 km/s. At this moment, a meteoroid exploded and destroyed. We use the meteor multicolour light curves revealed from a DSLR image in the RGB colour standard. We switch from the RGB colour system to Johnson's RVB colour system introducing colour corrections. This allows one to determine the colour characteristics of the meteor radiation. We are using a new approach based on colourimetry. Colourimetry of BGR three-beam light curves allows the identification of the brightest spectral lines. Our approach based on colourimetry allows direct measurements of temperature in the meteor trail. We find a part of the trajectory where the meteoroid radiates as an absolutely black body. The R/G and B/G light curves ratio allow one to identify the wavelengths of the emission lines using the transmission curves of the RGB filters. At the end of the trajectory, the meteoroid radiates in the lines Ca II H, K 393, 397 nm, Fe I 382, 405 nm, Mg I 517 nm, Na I 589 nm, as well as atmospheric O I 779 nm.

{\bf Key words:}\,\,space intrusions, methods: observational,
 processing techniques: photometric, objects: meteors

\end{abstract}

\section*{\sc introduction}
\indent \indent Total meteor luminous area includes both head and wake \cite{Popova01}. The precise boundary between the head and the wake of the luminous volume is not well known. In instantaneous meteor photographs ($6\cdot 10^{-4}$ s exposure) the average length of fast meteors is about 50 - 150 m \cite{Babadzanov}.

There are practically no direct measurements of temperature in the meteor trail. A model of a 1 cm-sized, -1 magnitude Leonid meteoroid with ablation of Mg atoms demonstrated a translational temperature field from 5000 K to 100000 K sized 4 meters across and tens of meters along the trajectory \cite{Pellinen}.

The best illustration of the problem is a comparison of the calculated and actual translational temperatures in the shock layer of the Apollo lunar return module. The calculated temperature behind the shock wave ($M$ = 32.5; $h$ = 53 km) is 58128 K. The actual temperature is about 11600 K \cite{Anderson06}.
Our approach based on colorimetry allow us direct measurements of temperature in the meteor trail.

The first observations of meteor spectra have been performed using binocular prism spectroscopes. It was found that the most prominent features in meteor spectra are emission lines. The two commonest and brightest lines were positively identified with the yellow neutral sodium doublet and the green neutral magnesium triplet \cite{Ceplecha98}.

Extensive spectroscopic programs were carried with a highly sensitive CCD video camera in BW mode \cite{Koukal}. The field of view is 60$^\circ$ x 48$^\circ$, the system uses a fast Tokina lens ($f/0.98$) with a variable focal length (3-8 mm). Resolution of the CCD chip enables the use of a holographic diffraction grating of 500 lines/mm. In this configuration, the spectrograph reaches a stellar limiting magnitude of +4.5$^{m}$. The faintest recorded meteors then have a relative magnitude up to +2.0$^{m}$. The magnitudes of meteors with a measurable spectrum have to be at least -2.0$^{m}$.

Below we consider the algorithm, focusing on a new approach based on colorimetry. We estimate the parameters of the intrusion based on the ablation model to estimate the characteristics of the meteoroid. We also compare the results of the theory and observations and present results and conclusions.

\section*{\sc observations}
\indent \indent 

Fig. 1 shows an image of Leonid meteor-6230 obtained by Mike Hankey in 2012. Observational data can be found from the archives of The American Meteor Society (AMS) \cite{Leonid}. 
Below we present the results of a photometric analysis of the meteor trail. Its complex structure is revealed from a DSLR image in the RGB color standard.
The time-frequency structure of the wake dynamics was calculated on the assumption that the duration of the meteor glow is 1 second. Spectral time analysis of the meteor light curves (Fig. 2) reveals two significant harmonics around 3 and 46 Hz.
The projection across the meteor track on the observer's plane shows a "wobbling" (movement in the direction perpendicular to the track). The main fluctuations in the brightness and wobbling of the meteor trail are observed at a frequency of about 3 Hz. The brightness variations in the integrated light are about 3\%. The amplitude of the wobbling is about 2\% (Fig. 3, 4).

\begin{figure}[!h] 
\centering
\begin{minipage}[t]{.45\linewidth}
\centering \epsfig{file = 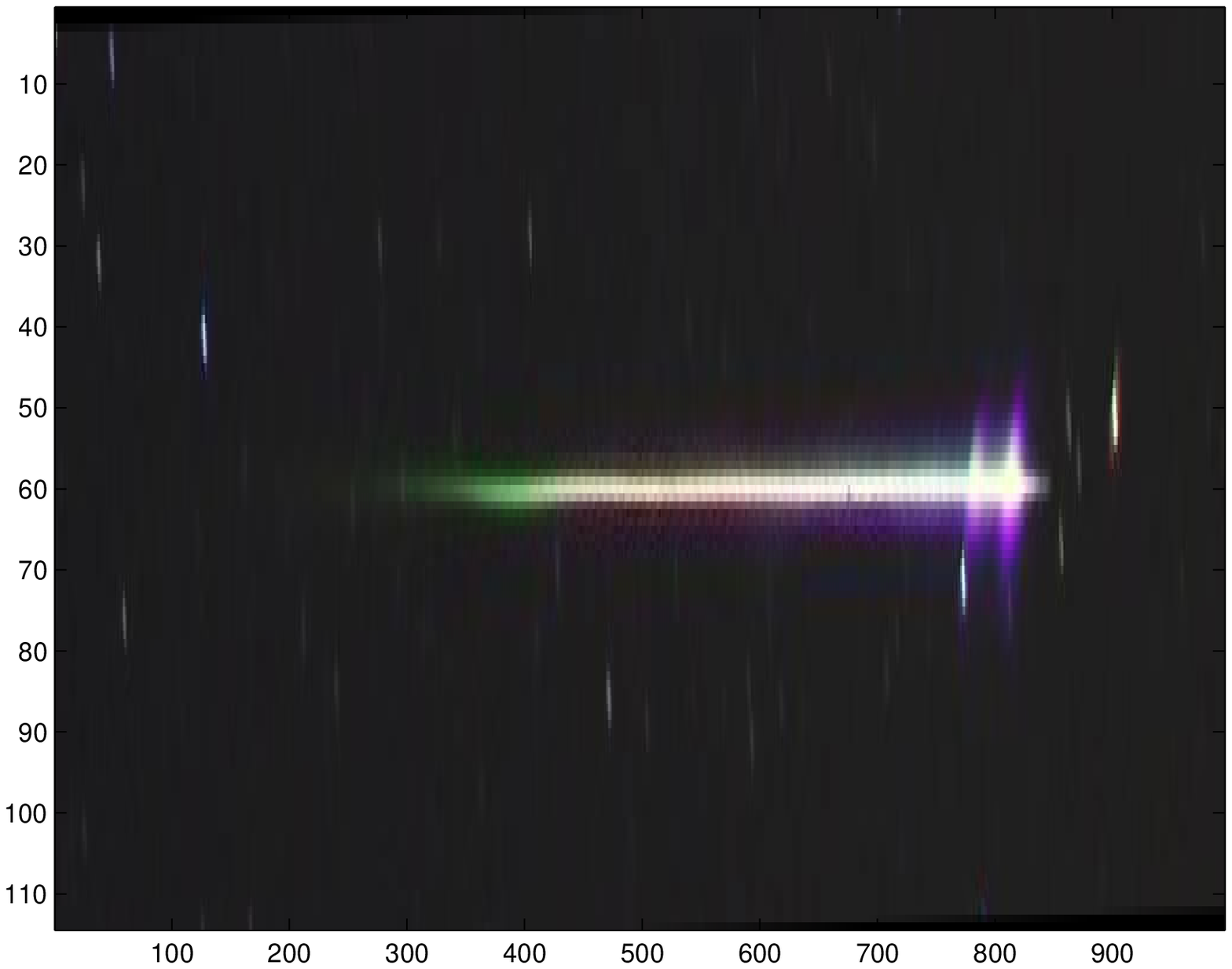,width =
1.0\linewidth} \caption{An image of Leonid meteor - 6230.}\label{fig1}
\end{minipage}
\hfill
\begin{minipage}[t]{.45\linewidth}
\centering
\epsfig{file = 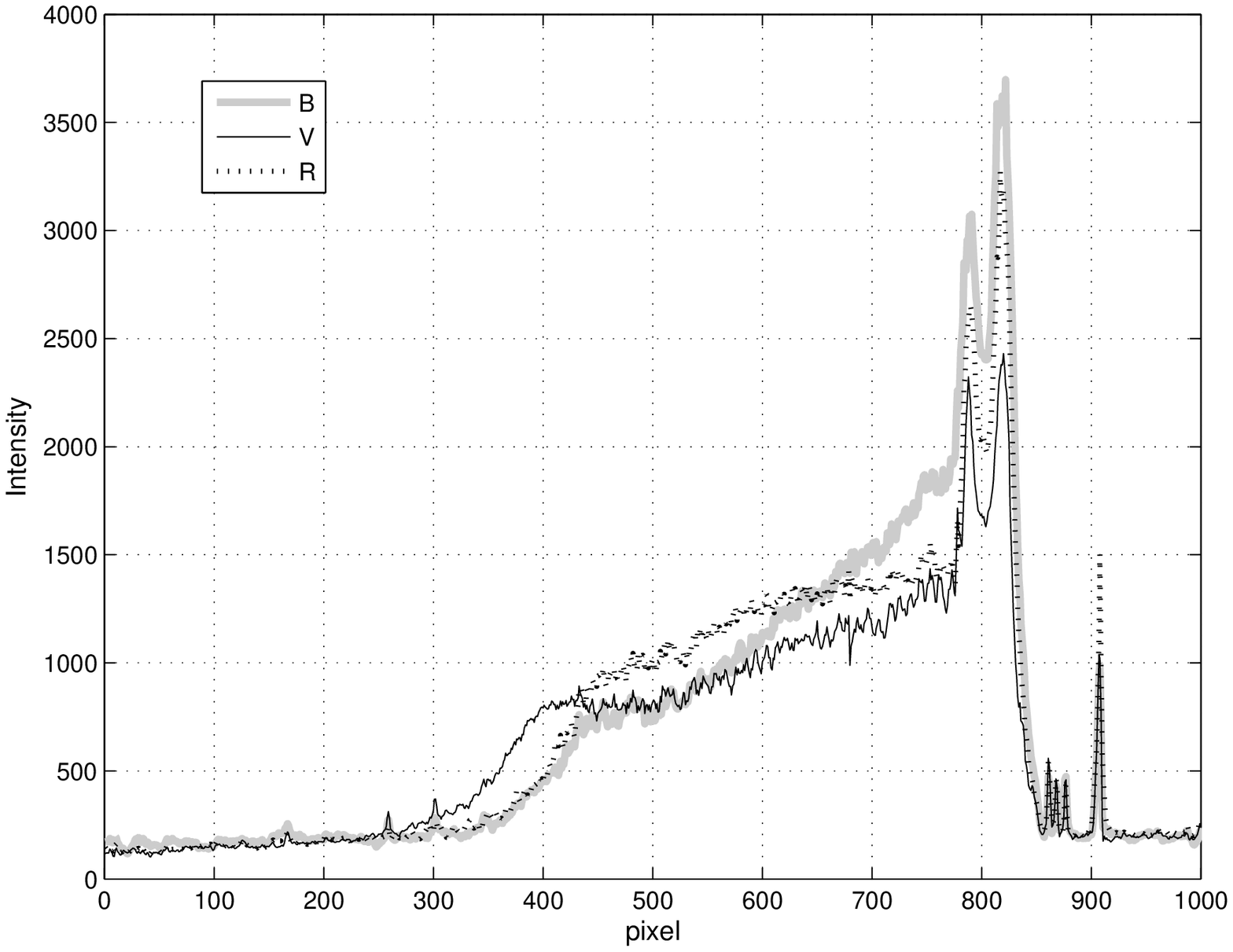,width = 1.0\linewidth}
\caption{The meteor light curves in BVR rays.}\label{fig2}
\end{minipage}
\end{figure}

\begin{figure}[!h] 
\centering
\begin{minipage}[t]{.45\linewidth}
\centering \epsfig{file = 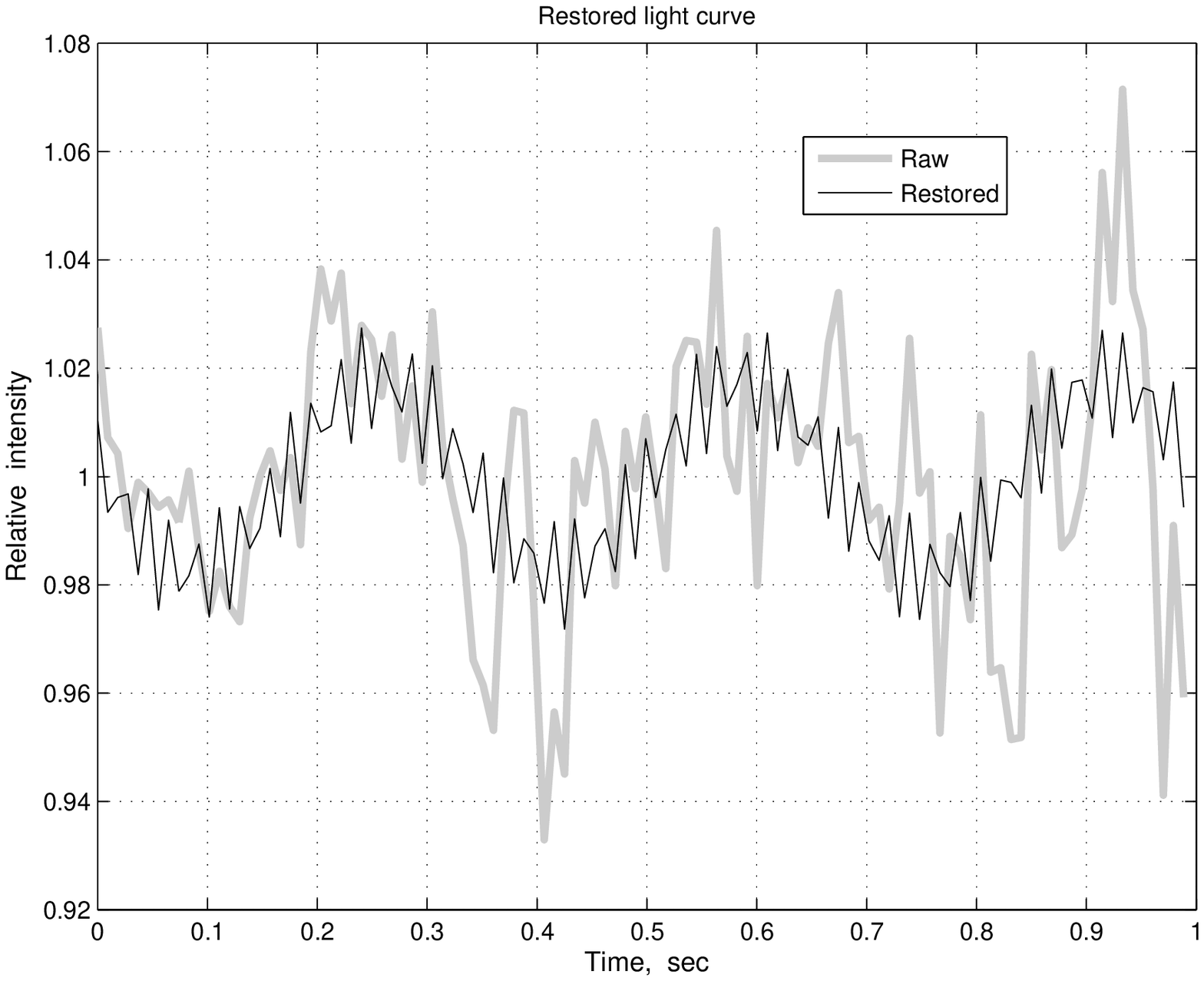,width =
1.0\linewidth} \caption{The brightness variations in the de-trended integrated light.}\label{fig1}
\end{minipage}
\hfill
\begin{minipage}[t]{.45\linewidth}
\centering
\epsfig{file = 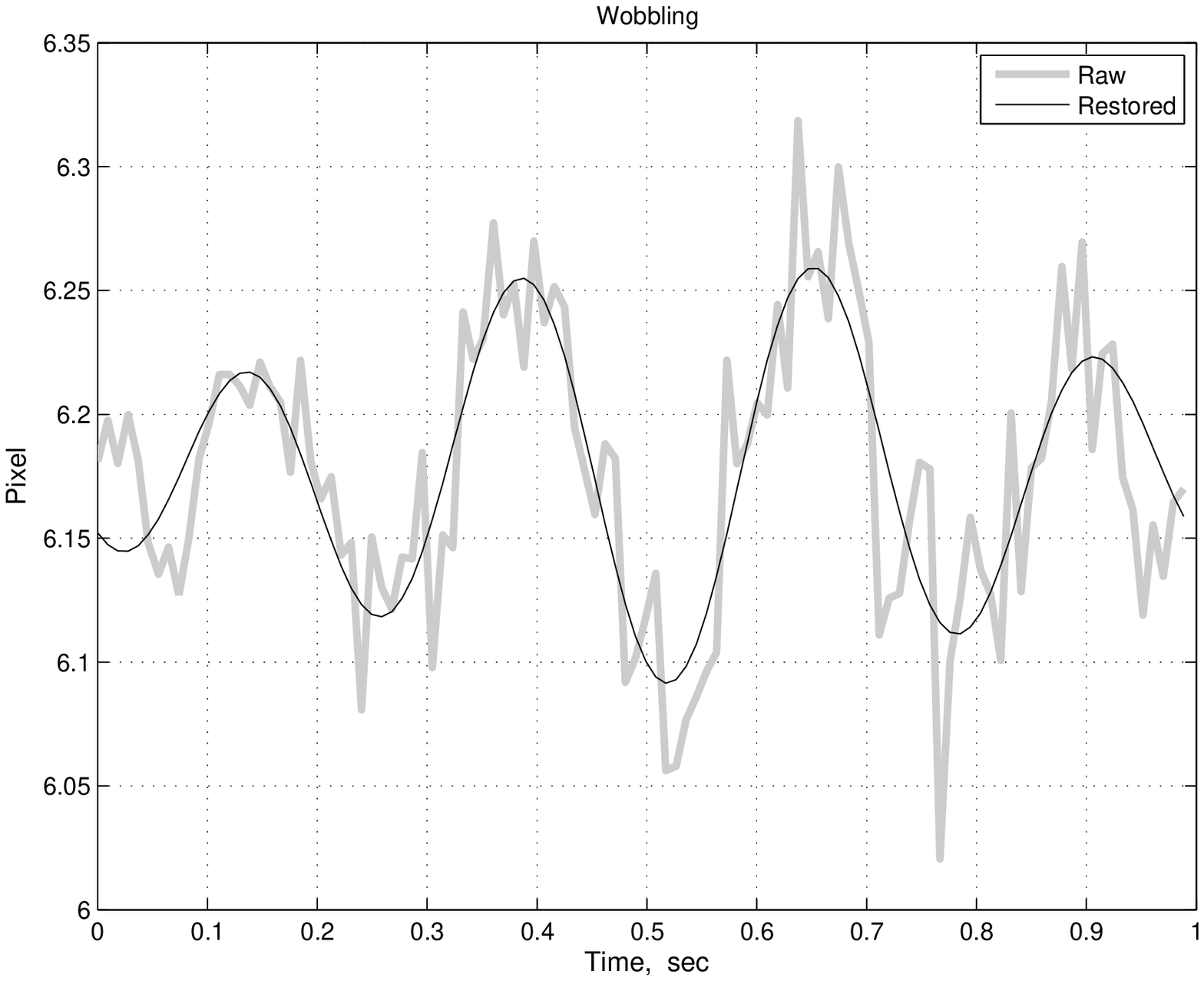,width = 1.0\linewidth}
\caption{The wobbling of the meteor trail.}\label{fig2}
\end{minipage}
\end{figure}

\section*{\sc analysis of the temporal features of the meteoroid trail}

\indent \indent Deviations from rectilinear motion indicate the presence of strength perpendicular to the direction of the flight path. This phenomenon is known as the Magnus effect.
It occurs when a stream of gas flows around a rotating body.
The equation of motion of a meteoroid in the linear approximation give the oscillation solutions \cite{Zhi20}.
The angular frequency of the brightness variations  $\Omega_{c}$ has the form:
\begin{equation}\label{}
\Omega_{c} = 2 \pi f_{c}= \sqrt{A} =
\sqrt{\frac{3}{4}\frac{C_{D}\,\rho \, \sin(\theta)\,
v^{2}}{r_{m}\rho_{m} H_{\rho}}}
\end{equation}
where $\rho $ is gas density, $\rho_{m}$ is the density of meteoroid, $\theta$ is the zenith angle, $C_{D}$ is the drag coefficient, $v$ is flow velocity, $H_{\rho}$ is the height of the homogeneous atmosphere, approximately 7 km, $r_{m}$ is the meteoroid radius.

The main frequency of rotation of Magnus $\Omega_{m}$ has the form:
\begin{equation}\label{}
 \Omega_{m}=2 \pi f_{m}=\frac{3}{2}\frac{\rho \,
\omega_{m}}{\rho_{m}}\cdot\sin( \theta)\cdot
(1+\frac{z}{H_{\rho}})
\end{equation}
where $\omega_{m}$ is the frequency of angular rotation.
We can obtain an estimate the angular frequency of the rotation of the meteoroid as \cite{Betz66}:
\begin{equation}\label{}
    \omega_{m}= \sqrt{\frac{C_{P}}{2}\, \rho \, S\, v^{3} / I \cdot T }
\end{equation}
Here $T$ is the time span, $S = \pi r_{m}^{2}$, $I=\frac{m}{2}\,{r_{m}^{2}}$, is the moment of inertia around the axis of rotation, $m$ is the mass. The power coefficient $ C_{P}$ is described by Betz's law \cite{Betz66} that indicates the power that can be extracted from the wind, independent of the design of a wind turbine. It has a maximum value $ C_{P}$ max = 16/27 = 0.593. 

\section*{\sc model of the ablation for estimation the meteoroid
characteristics}

\indent \indent We use the classical meteor ablation theory \cite{Downs98}, 
\cite{Campbel04}.
This model traces the variation of the meteoroid velocity,
deceleration, mass and visual radiation with time. By varying the
appropriate model parameters such as initial conditions, the
effect of them can be readily evaluated. The meteoroid momentum is
decreased through collisions with air molecules. The momentum
transfer can be quantified with the help of Newton's second law. 
A second equation, the differential mass equation, is derived
by assuming that the rate of loss of mass is proportional to the
kinetic energy transferred to the mass of air intercepted by the
meteoroid:
\begin{equation}\label{}
    dm/dt = (\Lambda \,A \,\rho \,v^{3} m^{2/3})/(2 \,\xi\, \rho_{m}^{2/3} )
\end{equation}
where $ \Lambda$ is the heat transfer coefficient, $ \xi$ is the
heat of ablation of the meteoroid.

The visual power radiated by the meteoroid is given by
\begin{equation}\label{}
    I=0.5 \,T_{I} \, dm / dt \, v^{2}
\end{equation}
where $ T_{I}$ is the luminous radiation efficiency. This may be
converted into a rough apparent visual magnitude for a
ground-based observer through the empirical relation
\begin{equation}\label{}
    V=6.8-1.086\, \ln I
\end{equation}

The table below lists typical initial values for the above
parameters used in the computer implementation of the model.

$\xi $  - meteoroid heat of ablation ($3\cdot 10^{6} $ J
kg$^{-1}$)

$ \Lambda$ - heat transfer coefficient  (0.15)

$ T_{I}$  - luminous efficiency $(1\cdot10^{-3})$.

The model use
for the density of the Earth's atmosphere a simple exponential
expression of the form $\rho = \rho_{0}\cdot \exp (-h/H)$.

The model described by the above equations can be implemented by
the Impact 4A Software \cite{Downs98}. The numerical integration of the
differential equations was performed using a Runge-Kutta 4
integrator. This was found to be stable at time steps of 0.001 s.

\section*{\sc  theory and observations}

\indent \indent How does the above theory correspond to reality?
Consider the light curve reproduced above (Fig. 2). The ablation
of a 900 kg solid meteoroid with an initial velocity of 36.5 km/s
was simulated using the Impact 4A Software considered above \cite{Downs98}.
The meteoroid reaches the height of 29 km with only 2\% of its
starting mass and exploded due to dynamical destruction. We show
that "tuning technique" in combination with a simulation model of
intrusion allows determining the meteor characteristics during
one-side observations.

Fig. 5 shows the speed of the meteoroid depending on the height,
the speed of rotation on the surface and the critical speed of
dynamic destruction. It is easy to see that the limit of dynamic fracture was reached at an altitude of about 29 km in a time of about 4.6 sec. At this
moment, a meteoroid exploded and destroyed. The brightness of the meteoroid reached
$V \simeq - 2 $ magnitudes. In this case, the residual mass was about
20 kg. Thus, ablation "ate" 98\% of the initial mass of the
meteoroid.

\begin{figure}[!h] 
\centering
\begin{minipage}[t]{.45\linewidth}
\centering \epsfig{file = 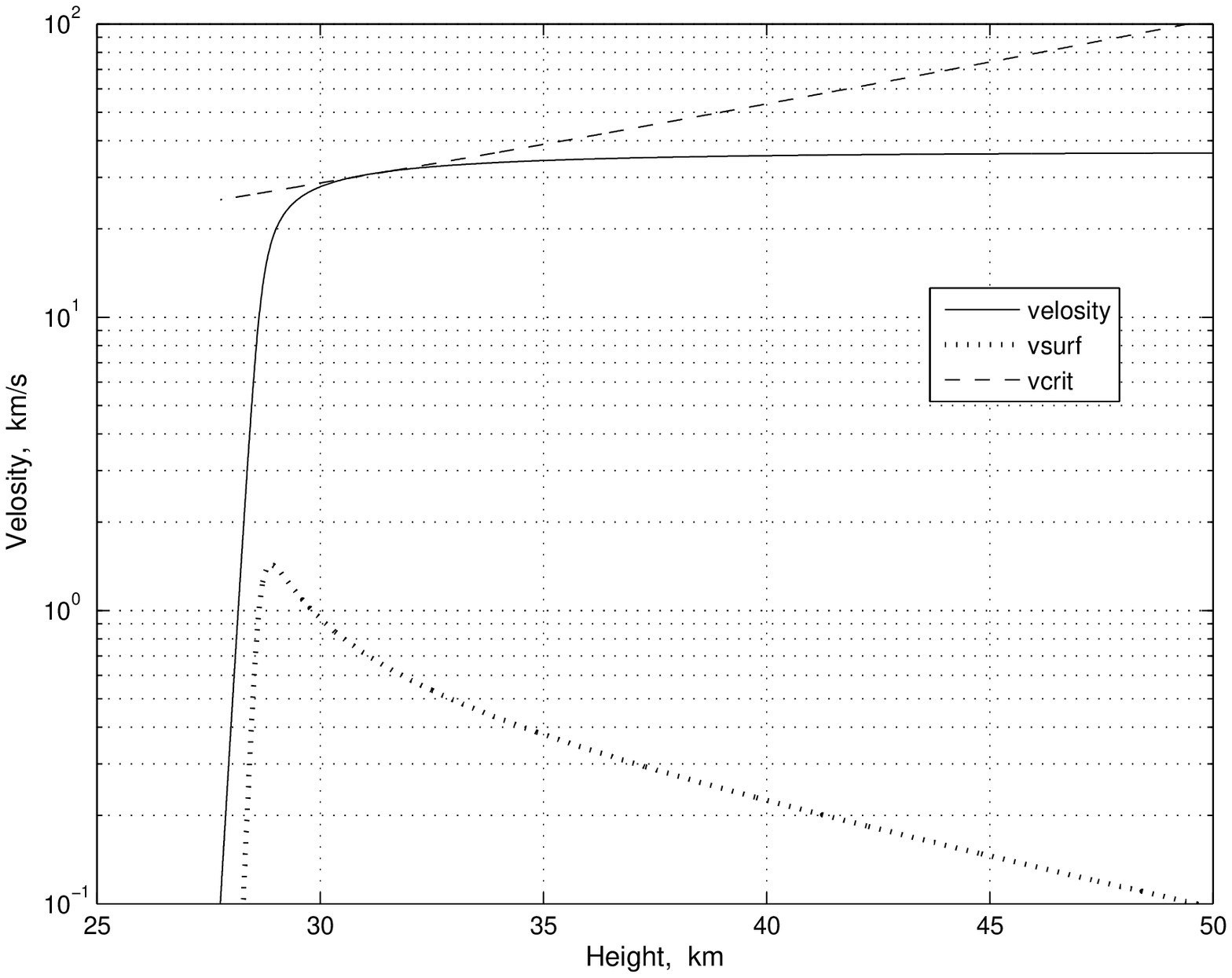,width = 1.0\linewidth}
\caption{The speed of the meteoroid depending on the height, the
speed of rotation on the surface and the critical speed of dynamic
destruction.}\label{fig1}
\end{minipage}
\hfill
\begin{minipage}[t]{.45\linewidth}
\centering
\epsfig{file = 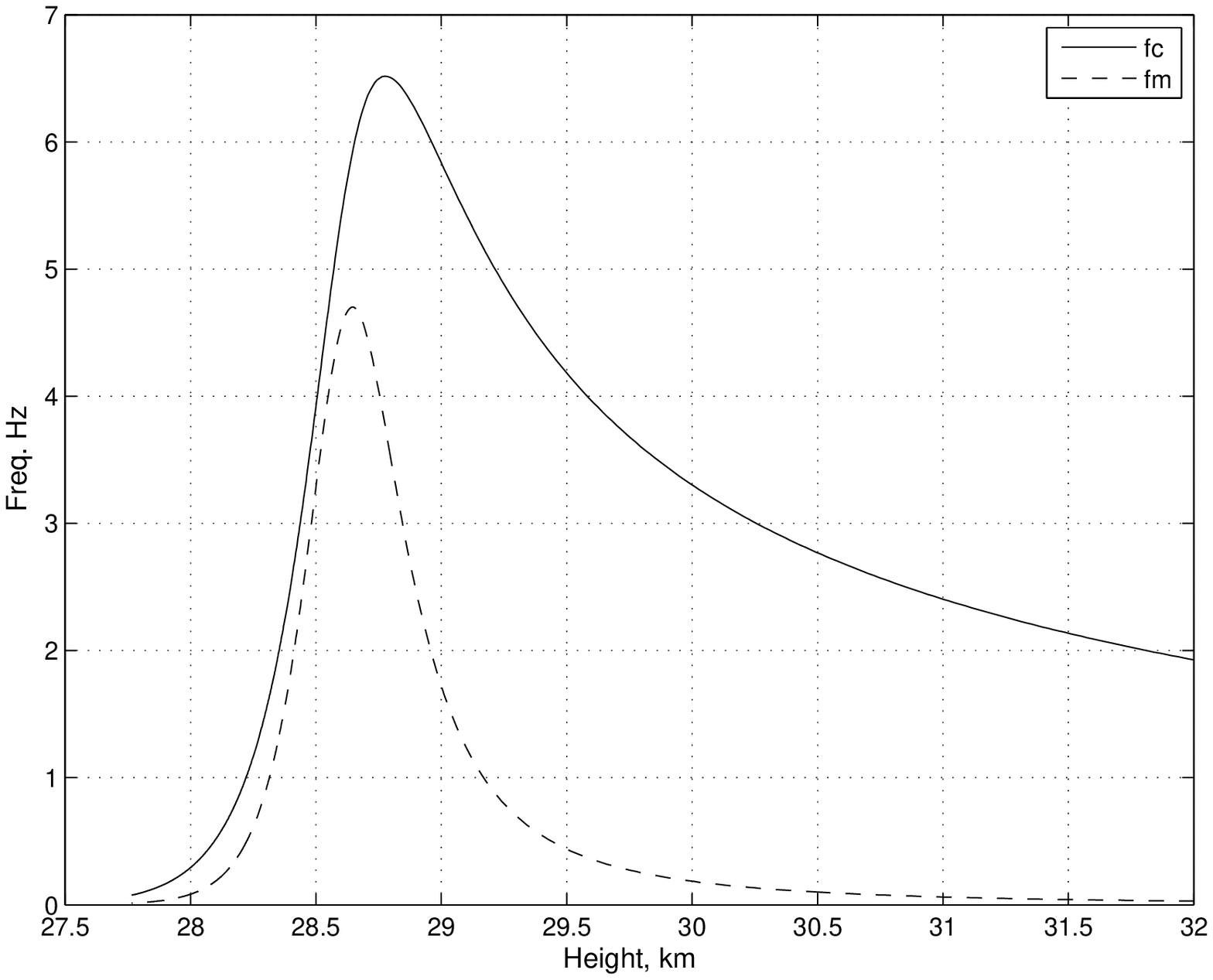,width = 1.0\linewidth} \caption{The values
of the frequencies $f_{c}$, $f_{m} $ depending on the
height.}\label{fig2}
\end{minipage}
\end{figure}

Fig 6 plays a crucial role in our work. The figure shows the
values of the frequencies $f_{c}$, $f_{m}$ depending on the
height. By varying the initial parameters of the model, the
primary mass and speed of the meteoroid at the boundary of the
Earth's atmosphere, we find the height at which the frequency
ratio $f_{c}/f_{m}$ coincides with the value found from the
measurements. Initial frequencies were calculated on the assumption that the duration of the meteor glow is 1 second. The final frequency values were determined on the time scale obtained as a result of the simulation.


\indent \indent This work is devoted to determining the meteor
characteristics during one-side observations. Photometry of the
meteor track indicates brightness variations, as well as
fluctuations in the trajectory transverse to the direction of
motion. 

Photometry of the meteor made it possible to determine the
oscillation frequency estimated on a time scale of about one sec, as well as the transverse oscillations of the trajectory.
We use "tuning technique" in combination with a simulation model
of intrusion for determining the meteor characteristics. It was
found that the progenitor of the meteor is an object weighing 900 kg at a speed of 36.5 km/s, The meteoroid reached a critical value of
the pressure at an altitude of about 29 km in a time of about 4.6 sec with a residual
mass of about 20 kg, and a residual speed of about 28 km/s. At this
moment, a meteoroid exploded and destroyed.

\section*{\sc meteor colorimetry with CMOS cameras}

\indent \indent The colorimetry of celestial sources in astronomy is based on Johnson's UBVR color system. Color diagrams U - B, B - V, V - R, allowing to determine the effective temperature of radiation sources, as well as the characteristics of hydrogen plasma. In the case of meteors, colorimetry can determine the effective temperature in the case of optically thick plasma. Colorimetry of BGR three-beam light curves, allowing identification of the brightest spectral lines. We can use the transmission curves of the filters and the balance of the intensity of the BGR light curves, as shown in Fig. 9. Thus, although colorimetry is not a universal tool, it can help diagnose meteoric plasma.

When observing meteors, color CMOS cameras are widely used, in which the Adobe RGB (1998) color system is implemented \cite{Adobe}. Colors are indicated as triplets [R, G, B], where each of the R, G, and B components has amplitude values ranging from 0 to 1. When displaying color, the exact chromaticity values of the reference white point have amplitudes of [1,1,1], reference black point [0,0,0] and primaries [1,0,0], [0,1,0] and [0,0,1]. Varying the amplitude of the primary colors the RGB colors systems can be adjust at a predetermined color temperature of the blackbody radiation intensity distribution determined by the Planck curve.

Centers and equivalent widths of RVB bands of the Johnson system are [700, 550, 440] nm and [220, 90, 98] nm, Adobe RGB [685, 545, 465] nm, and [110, 50, 50] nm (without response in the near infrared), respectively \cite{Adobe}, \cite{Allen}. Therefore, to switch from RGB color system to   Johnson's RVB color system, it is necessary to introduce color corrections.

In UBVR color systems the Johnson B - V , V - R color diagrams attached to the scale of the effective temperature blackbody radiation and the main sequence stars in the range of 2650 K to 50000 K \cite{Straizys}, \cite{Allen}. Knowing the color corrections, one can easily switch from the RGB color system of a CMOS camera to the RVB   system of Johnson and determine the color characteristics of the meteor radiation.

The color correction is usually operate directly with the values of the intensities of red, green, and blue channels of the RGB . Most meteors images from the archives of the AMS, the IMC set the color balance D65, which corresponds to the light scattering of the clear daylight sky (the Daylight). The color temperature of the scattered solar radiation corresponds to 5770 K \cite{Williams}. 

Using the data for the BVR bands of the Johnson and Adobe BGR systems presented above, for the Planck curve with a solar radiation temperature of 5770 K, we obtain the band intensity ratios

\begin{eqnarray*}
  FB_{J}= 1.88 * FB_{Adobe} \\
  FV_{J}= 1.79 * FG_{Adobe} \\
  FR_{J}= 1.93 * FR_{Adobe} \\
\end{eqnarray*}  

Hence, the color corrections for the color index B - V is - 0.05 and V - R is +0.08 magnitudes. Taking into account the color of the solar radiation index (B - V)$_{\bigodot}$ = 0.653 $\pm$ 0.003, (V - R)$_{\bigodot}$ = 0.356 $\pm$ 0.003 \cite{Williams}, we obtain the final values of the corrections:

\begin{eqnarray*}
(B - V)_{J} = (B - G)_{Adobe} +0.60 \\
(V-R)_{J} = (G - R)_{Adobe} +0.44 \\
\end{eqnarray*}  


In some cases, when shooting on cameras, the color temperature for daylight is automatically set to 5200 K. In this case, the color corrections for the color indexes will be:

\begin{eqnarray*}
(B - V)_{J} = (B - G)_{Adobe} +0.71 \\
(V-R)_{J} = (G - R)_{Adobe} +0.71 \\
\end{eqnarray*}  


Thus, having the values of the color corrections, one can easily switch from the RGB color system of a CMOS camera to the RVB   system of Johnson and determine the color characteristics of the meteor radiation.

Colorimetry is a quantitative method for analyzing radiation. It allows one to diagnose the radiation of the emitting plasma based on theoretical diagnostic color diagrams calculated for various radiation sources. To develop two-color diagrams, we used the color indices of absolute blackbody radiation \cite{Straizys}. At first, in each filter the intensities $FR, FV, FB$ of a meteor were calculated in Fig. 2. 
Based on the obtained data, the color indices of the intrinsic emission were determined:
\begin{equation}\label{}
    V-R=-2.5\lg\left[\frac{FV}{FR}\right]+\Delta(VR)
\end{equation}
\begin{equation}\label{}
    B-V=-2.5\lg\left[\frac{FB}{FV}\right]+\Delta(BV)
\end{equation}
where $\Delta(VR)$, $\Delta(BV)$ are normalization coefficients. These coefficients take into account the color corrections. 

Fig. 7 shows the temporal evolution of the color characteristics of a meteor over the whole period of time. 
Over some period of time, the meteor radiation is localized in an area close to the radiation of a completely black body (Fig. 8). Absolute blackbody radiation is indicated by the dash line.

The main part of the meteor track shows a blackbody spectrum with temperatures varying from 4200 K at the beginning to 5500 K at the end of the path. Fig. 2 shows several flares on the meteoroid light curves. At the beginning (during the flash) and at the end (during the explosion), the color tracks  move in the direction occupied by optically thin plasma, the line spectrum prevails.
Note that knowing the blackbody temperature, height, and luminosity at a track point, it is easy to determine its cross-sectional area.

\begin{figure}[!h] 
\centering
\begin{minipage}[t]{.45\linewidth}
\centering \epsfig{file = 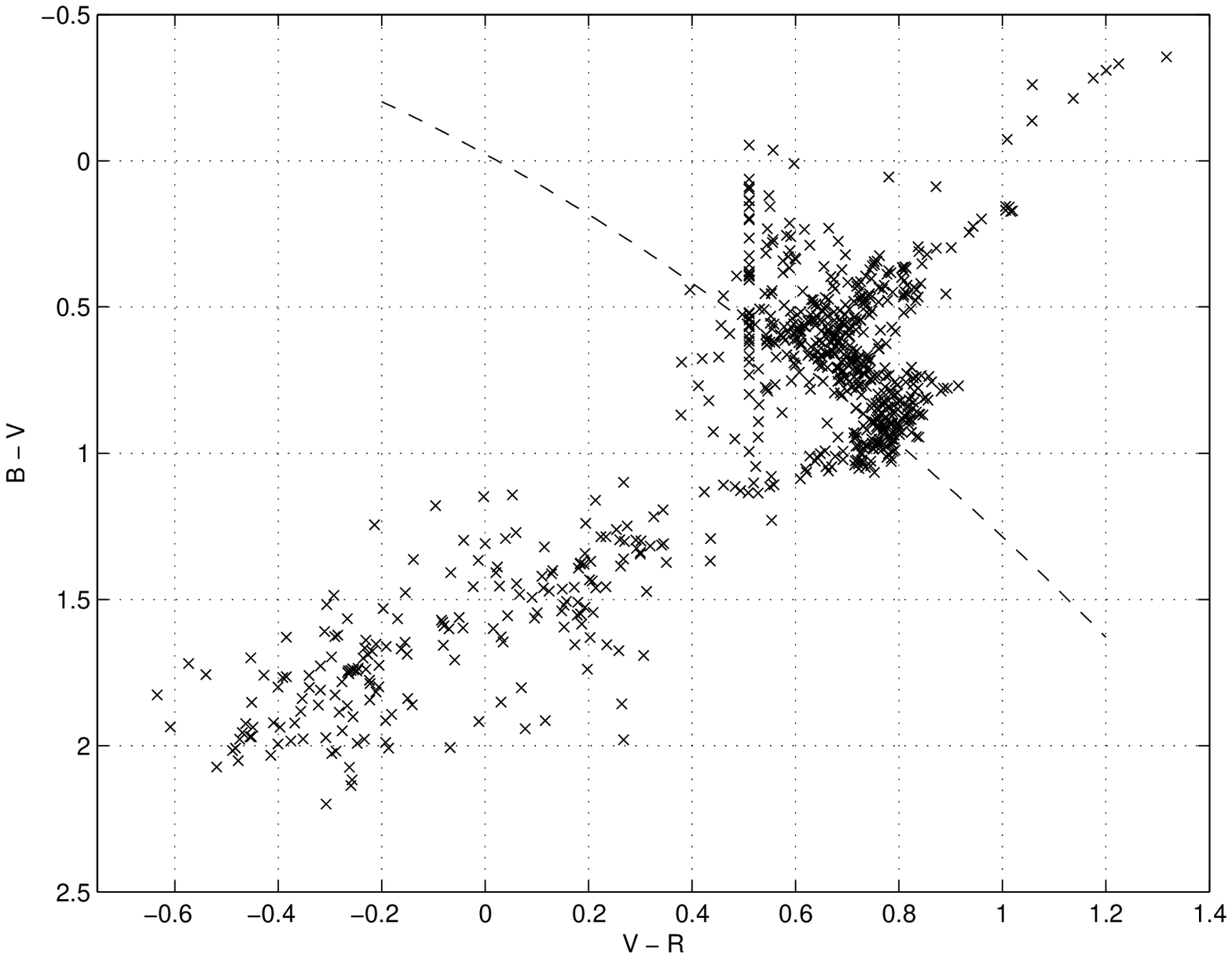,width =
1.0\linewidth} \caption{The color characteristics of a meteor over the whole period of time. Blackbody radiation is indicated by the dash line.}\label{fig1}
\end{minipage}
\hfill
\begin{minipage}[t]{.45\linewidth}
\centering
\epsfig{file = 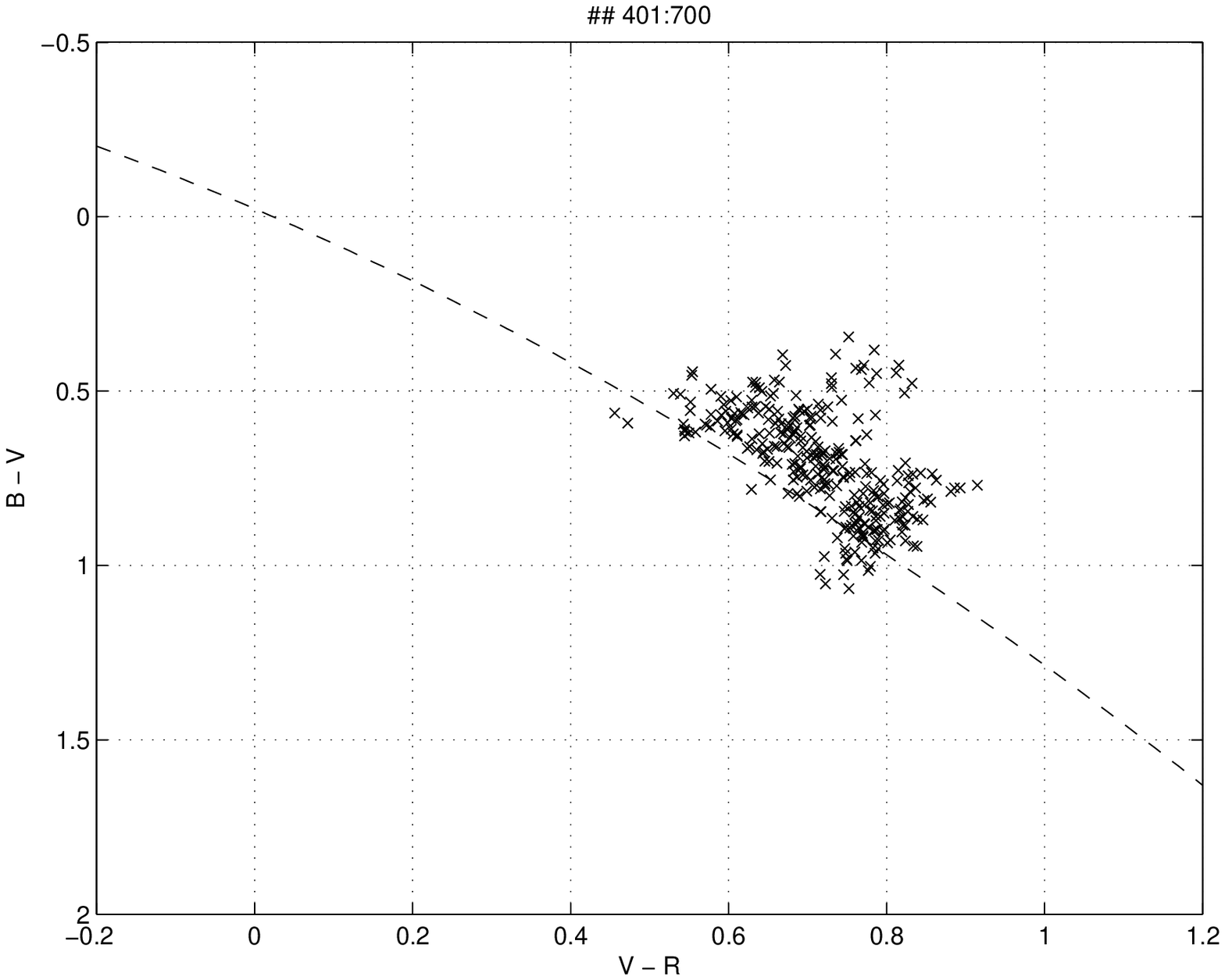,width = 1.0\linewidth}
\caption{Variations in the meteor brightness indicate changes in the temperature in the range 4200 - 5500 K. }\label{fig2}
\end{minipage}
\end{figure}

\begin{figure}[!h] 
\centering
\begin{minipage}[t]{.45\linewidth}
\centering \epsfig{file = 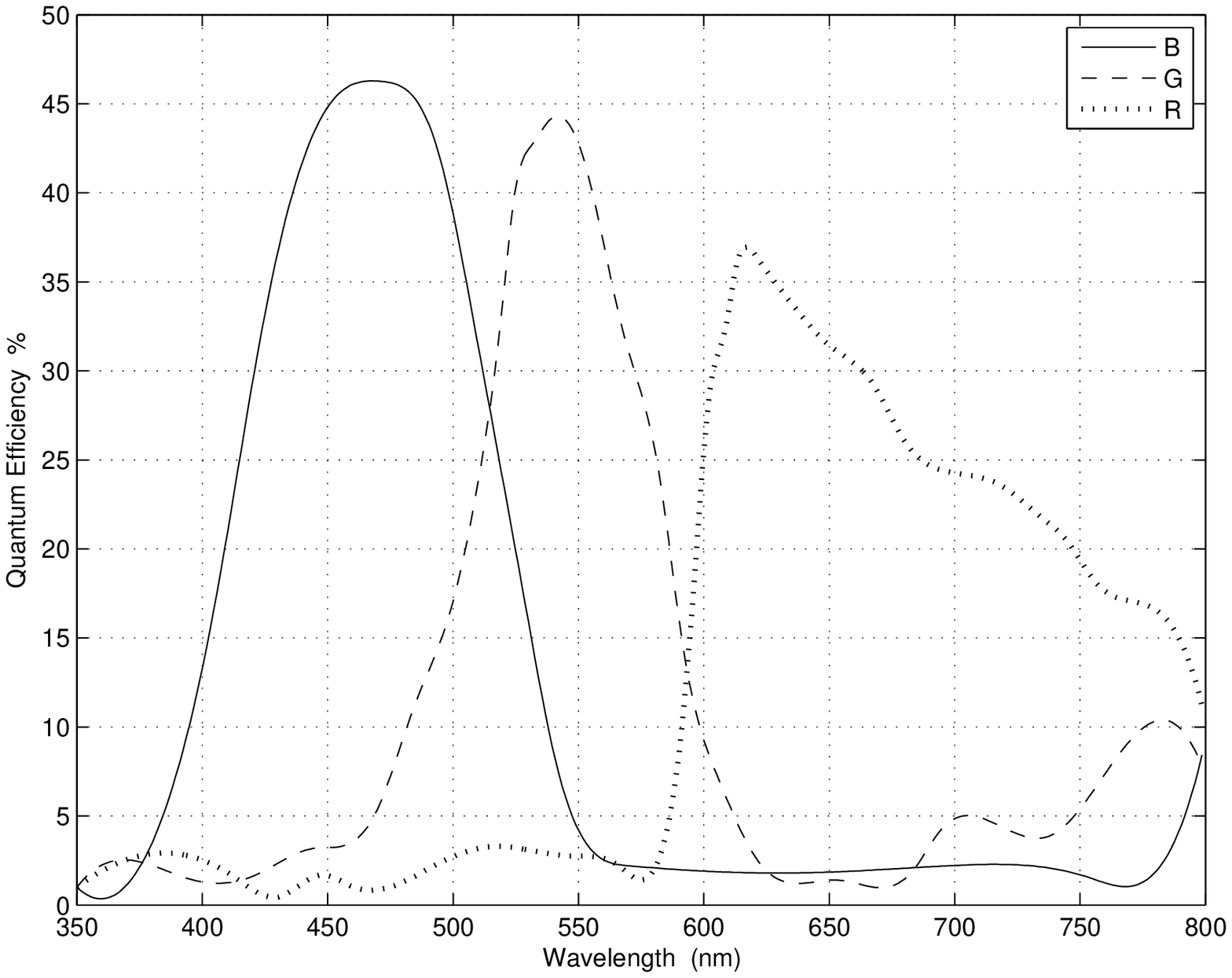,width =
1.0\linewidth} \caption{The BGR spectral response curves.}\label{fig1}
\end{minipage}
\hfill
\begin{minipage}[t]{.45\linewidth}
\centering
\epsfig{file = 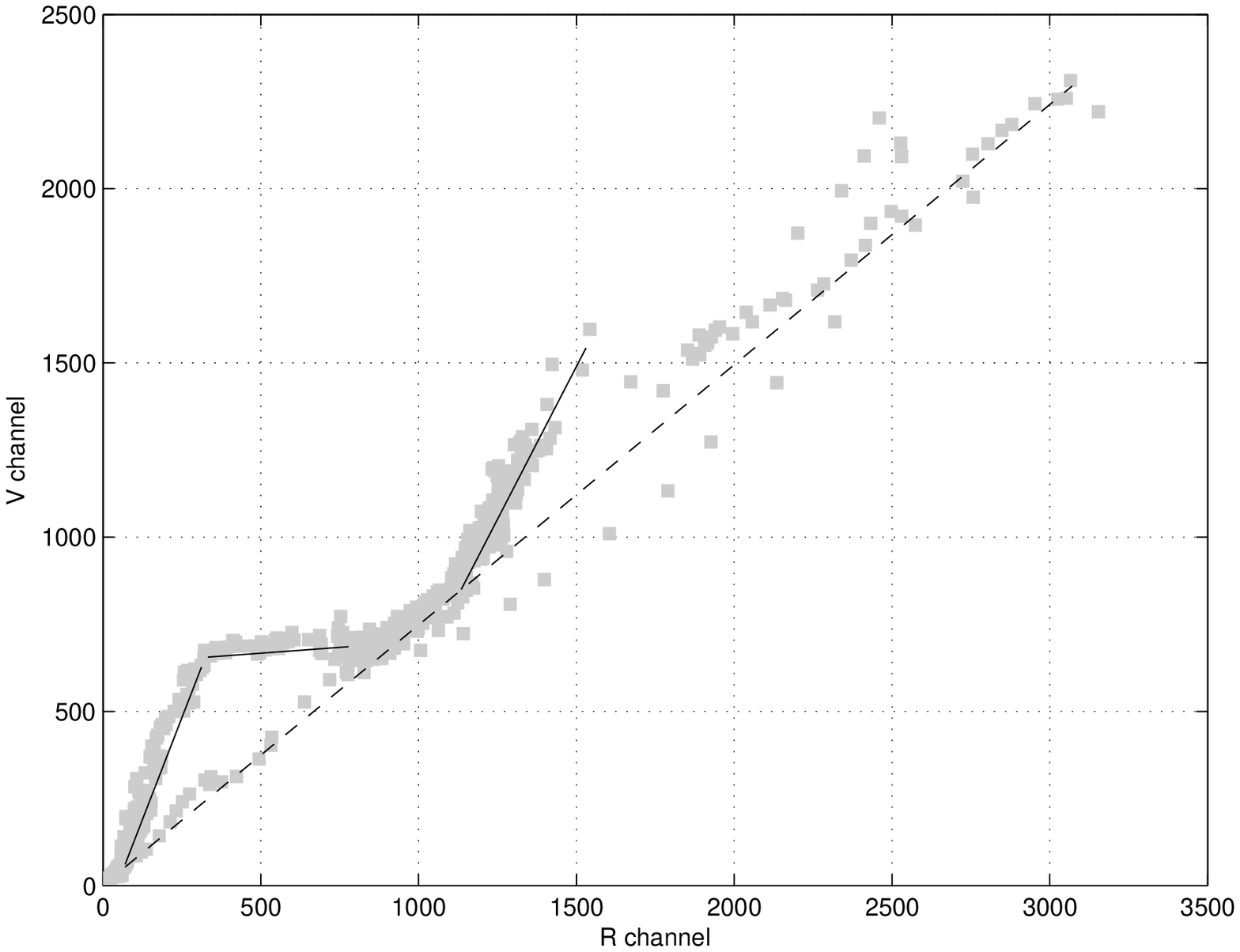,width = 1.0\linewidth}
\caption{The V/ R (V vs R) diagram.}\label{fig2}
\end{minipage}
\end{figure}

\section*{\sc Line identifications}

\indent \indent The V/R and B/V light curves ratio allow one to identify the wavelengths of the emission lines using the transmission curves of the RGB filters. Several bright lines have been successfully identified using the colourimetry method. Fig. 9 displays RGB sensor sensitivity in terms of absolute Quantum Efficiency (QE) \cite{PointGray}. QE is the percentage of photons converted to electrons at a specific wavelength, expressed in percentage.

To understand the essence of the line identification method, let's imagine that we have one spectral line. Since the transmission curves of RGB filters include common parts of the spectrum, the line will respond in two filters. The position of the line in the spectrum will depend on the ratio of the quantum efficiencies of the filters at the wavelength of the spectral line. It is easy to see that the method will work for multiple spectral lines as well.

The diagram in Fig. 10 shows the result of dividing the V- and R-light curves. V / R (V versus R) provide information about radiation sources. The portions of straight lines on the graph allow us to identify spectral lines. In particular, the cotangent of the line segment angle dV/dR determines the ordinate of a straight line, the intersection of which with the graph V / R functions determines the values of the wavelengths of the spectral lines (Fig. 11, 12).  The diagrams also show the lines responsible for the emission of atmospheric oxygen O I (779 nm) and nitrogen N2 (631 nm) in the meteor track and the high-temperature component of the spectrum, emitting as a black body.

At the beginning of the trajectory (dB/dV = 1.38), the emission of magnesium Mg I 518 nm, nitrogen N2 631 nm, iron Fe I 373, 649 nm, and the Balmer hydrogen $H_{\alpha}$ line 656 nm is observed (Fig. 12).

At the end of the trajectory, a global explosion and complete destruction of the meteor is observed. The cloud shone in the lines Ca II H, K 393, 397 nm, Fe I 382, 405 nm, Mg I 517 nm, Na I 589 nm, as well as atmospheric O I 779 nm (Fig. 11, 12, 13).

Knowing the RGB light curves, the QE values of the filters at the emission line wavelengths, it is possible to estimate the intensity of the lines in the spectrum (Fig. 13).

\begin{figure}[!h] 
\centering
\begin{minipage}[t]{.45\linewidth}
\centering \epsfig{file = 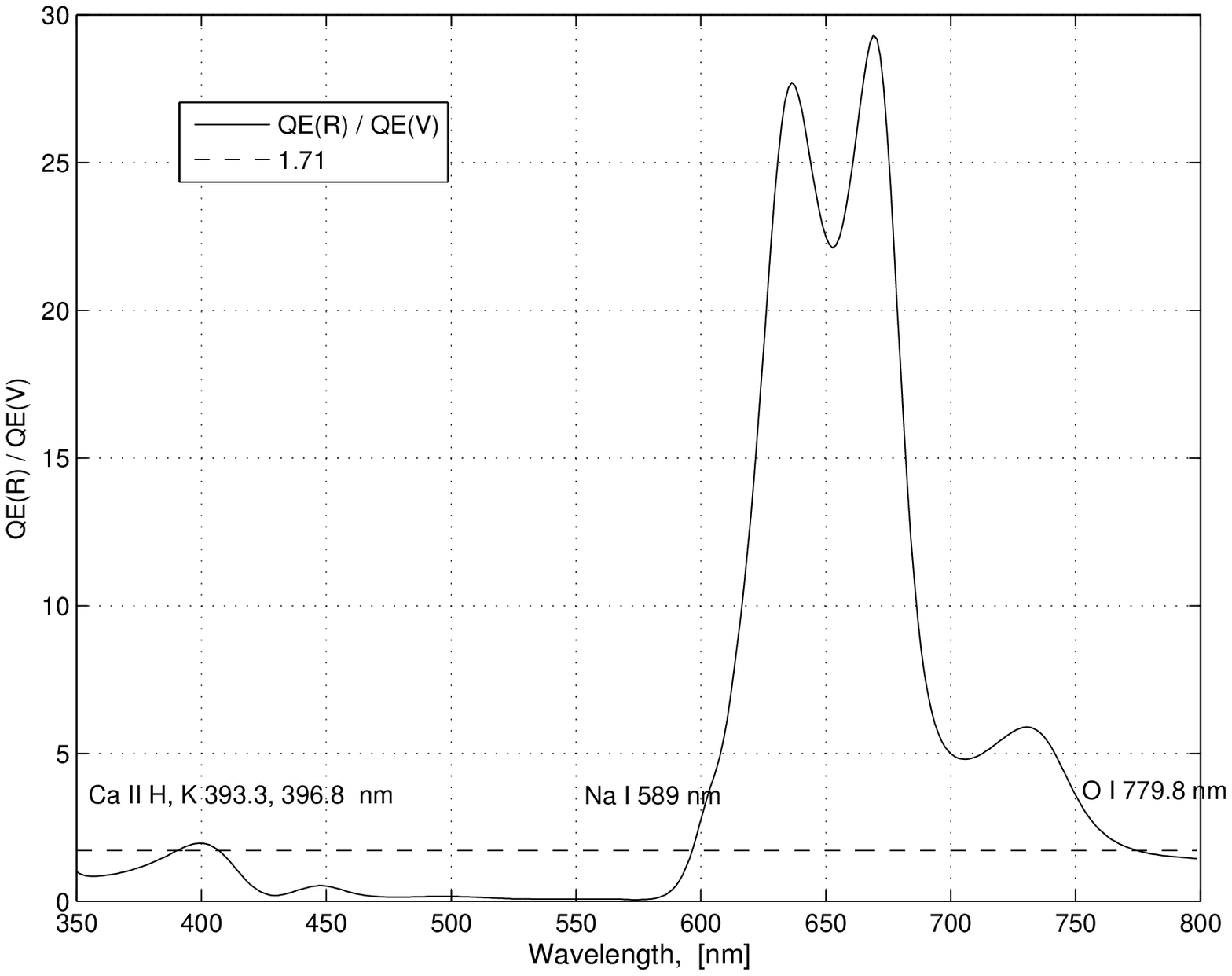,width = 1.0\linewidth}
\caption{The V / R diagram.}\label{fig1}
\end{minipage}
\hfill
\begin{minipage}[t]{.45\linewidth}
\centering
\epsfig{file = 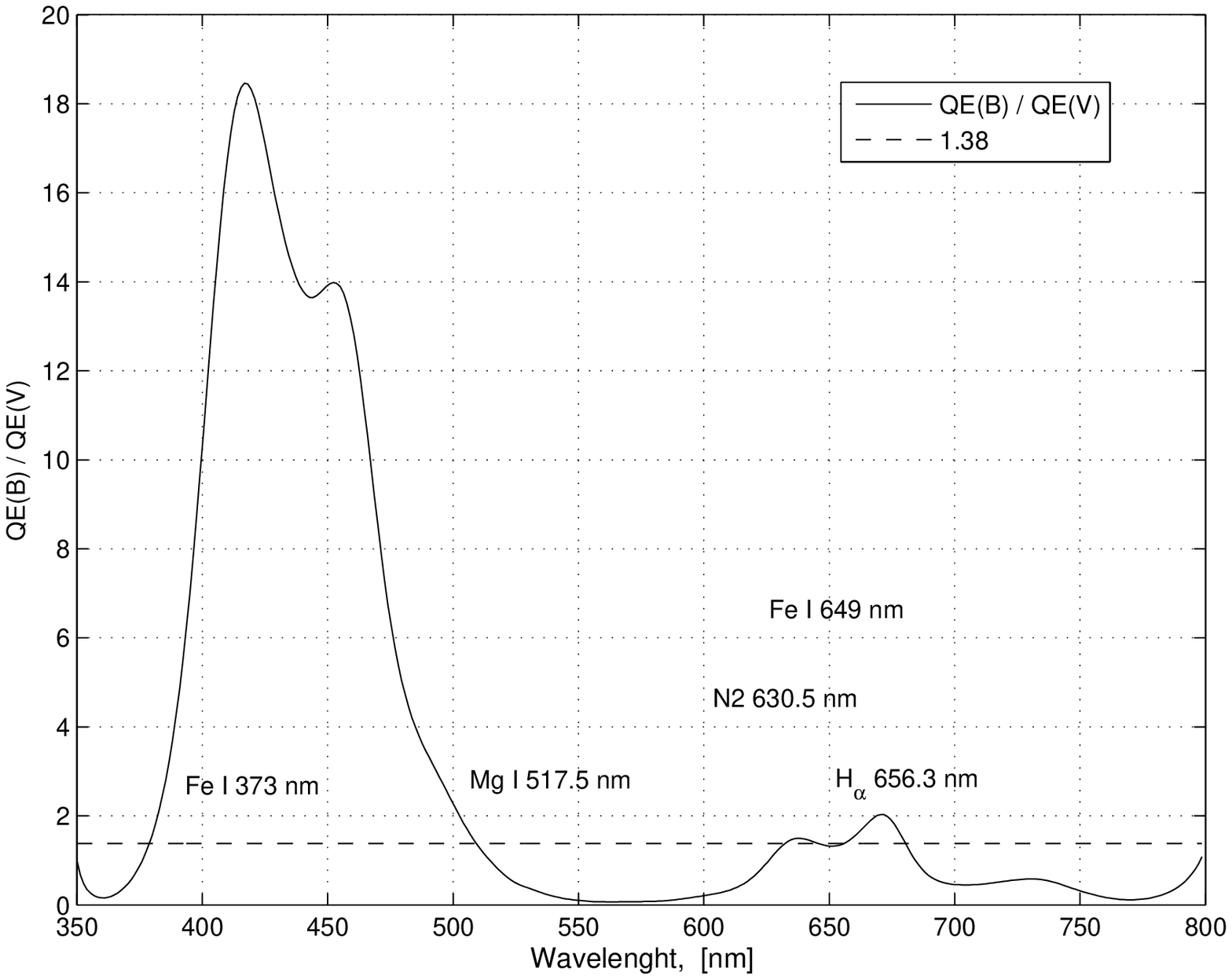,width = 1.0\linewidth} \caption{The B / V diagram.}\label{fig2}
\end{minipage}
\end{figure}

In real cases, both a continuous spectrum and a line radiation are recorded. If we take into account only self-absorption in the line under the condition of local thermodynamic equilibrium, then the intensity of radiation in the line $I(\nu)$ has the form:

\begin{equation}\label{}
 I(\nu)=B_{\nu}(T)\cdot\frac{\pi e^{2}f_{mn}}{mc\triangle v_{L}}\left\lfloor 1-exp\left ( -\frac{h\nu}{kT}\right)\right]LN_{n}
\end{equation}
where $B_{\nu} (T)$ is the Planck intensity at temperature $T$, $e$ and $m$ are the electron charge and mass, $f_ {mn}$ is the oscillator force, $\triangle v_ {L}$ is the line width, $h$ is the Planck constant, $k$ is Boltzmann's constant, $LN_ {n}$ is the number of particles per unit area.

\begin{figure}[!h] 
%
\centering \epsfig{file = 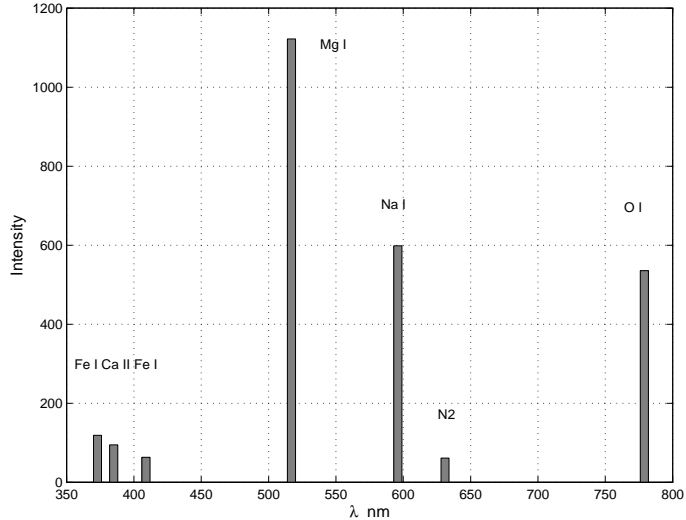,width = 0.5\linewidth} \caption{The emission spectrum of cloud blast with atmospheric emissions of OI and N2.}\label{fig1}
\end{figure}

\section*{\sc Air blast}

\indent \indent At the end of the light curves the meteoroid reached a critical value of the pressure on the surface that exceed the strength of the material. At this moment, a meteoroid exploded and destroyed. To estimate the characteristics of the explosion cloud, we will use the theory of a strong explosion in the atmosphere \cite{Landau}, \cite{Zeld}. As shown in \cite{Landau}, the gas movement, in this case, will be self-similar. The radius $R$ of the sphere of the shock wave depending on the explosion energy $E$, the density of the atmosphere $\rho$ and time has the form:
\begin{eqnarray*}
R \simeq \left( \frac{E}{\rho}\right)^{1/5} t^{2/5}\\
\end{eqnarray*}  

The self-similar nature of the motion is violated when the pressure at the front of the blast wave reaches the atmospheric value $p$. In this case, the radius of the sphere is determined by the expression:
\begin{eqnarray*}
R \simeq \left( \frac{E}{p}\right)^{1/3}\\
\end{eqnarray*}  

The simulation results allow us to estimate the parameters of the meteor during the explosion: $v$ = 28 km/s, $m \simeq 20$ kg, $E =10^{10}$ J, air pressure - 12 mbar, velocity sound - 315 m/s. Thus it follows: the radius of the explosion cloud is about 200 m, the cloud formation time is 0.64 sec. The optical depth of nitrogen in the cloud, depending on the temperature of the meteoroid at the moment of the explosion, was, according to (9), 0.1 at 4200 K, 1.0 at 4600 K, and more than 60 at 5500 K.

\section*{\sc Conclusion}

\indent \indent Our colorimetric approach to quantifying the characteristics of meteors allows us to estimate the temperature, chemical composition, and spectra of meteors containing emission lines. By observing meteors with color CMOS cameras, we can switch from the Adobe RGB color system to the Johnson RVB color system using color correction and determine the color characteristics of the meteor radiation. We analyze an image of Leonid meteor-6230 obtained by Mike Hankey in 2012. 

On the trajectory segment, the meteoroid is emitted by a completely black body. The temperature at first is about 4200 K, and at the end - about 5500 K. At the beginning and end of the trajectory, the emission of sodium Na I 589 nm and magnesium Mg I 517 nm, as well as the lines Ca II H, K 393, 397 nm, Fe I 382, 405 nm, is observed.

In addition to blackbody radiation, atmospheric oxygen radiation O I 615 nm and nitrogen
N2 631 nm is observed. This radiation can be seen in the live track image as an optically thin component.


\end{document}